# The substitutional atomic distance model for predicting lattice thermal conductivity in alloys

Zhicheng Zong[1,2], Tianhao Li[1,2], Shixian Liu[3], Haisheng Fang[1], Nuo Yang[2*]

1. School of Energy and Power Engineering, Huazhong University of Science and Technology, Wuhan 430074, China.
2. College of Science, National University of Defense Technology, Changsha 410073, China.
3. Independent Researcher, Xinzhou 034000, China.

*Corresponding email: nuo@nudt.edu.cn (N. Yang)

**Abstract**

Understanding phonon transport in alloys is crucial for the design of high-performance electronic and thermoelectric devices. However, conventional theoretical models fail to provide a clear physical picture of phonon scattering caused by atomic disorder in alloys, and their prediction accuracy is limited. In this work, a new substitutional atomic distance model for alloys is proposed, providing an intuitive physical picture. SiGe and InGaAs alloys are taken as representative systems, and their thermal conductivities are calculated, showing good agreement with previous experimental measurements. The results indicate that alloy scattering plays a dominant role in reducing thermal conductivity. This study provides new insights into phonon transport in alloys and offers guidance for tailoring thermal properties through compositional engineering.

## 1. Introduction

Alloys are widely utilized in modern electronic devices[1,2], because physical properties can be tunable by compositional variation[3,4]. By adjusting the composition, key properties such as band structure and thermal transport can be effectively engineered to meet specific application requirements. However, such compositional variation inevitably introduces atomic-scale disorder through the random distribution of different atomic species. This intrinsic disorder plays a critical role in determining carrier and phonon transport behaviors.

Among the various physical properties, thermal conductivity of alloys plays a central role in heat dissipation and device reliability. It critically influences device performance, particularly in nanoscale and high-power applications. However, accurately predicting the thermal conductivity remains challenging due to the complex interplay between disorder-induced scattering and phonon transport, as well as the lack of a clear spatial description of scattering processes.

Various approaches have been developed to address this problem, including boltzmann transport equation (BTE) calculations, molecular dynamics (MD) simulations, and analytical or semi-analytical theoretical models. Although BTE and MD simulations are widely used, they face considerable challenges in accurately describing strongly disordered systems, largely due to computational limitations. For BTE methods, one common strategy is to employ special quasirandom structures (SQS)[5] to model alloy disorder, but such calculations are severely constrained by the maximum feasible supercell size. In contrast, MD simulations are more fundamentally limited by the accuracy of interatomic potentials; the intrinsic randomness and local lattice distortions in alloys make thermal conductivity predictions highly sensitive to potential reliability.

In theoretical studies of phonon thermal conductivity in alloys, the Virtual Crystal Approximation (VCA)[6-8] was initially developed as a fundamental framework. Subsequent studies have widely employed the VCA approach[9], and its validity in describing phonon dispersion of alloys has been supported by calculations using SQS[5] reported by Marco et al[10]. The main challenge lies in the accurate description of alloy disorder scattering.

While existing theoretical models fail to provide a transparent physical picture of alloy disorder scattering processes. The Klemens model[11,12], a classic analytical method based on perturbation theory and the Debye approximation, provides a simple phonon scattering rate, yet suffers from low quantitative accuracy due to its simplified assumptions. The Tamura model[13,14] further improved the formalism by abandoning the Debye approximation and enabling mode-resolved scattering calculations using first-principles phonon dispersions and eigenvectors within the VCA framework, achieving better accuracy while still focusing mainly on mass disorder. Subsequent models have introduced force-constant disorder on this basis, yet current approaches still lack a physically transparent understanding of alloy phonon scattering rates. These limitations motivate the development of a theoretical approach that is both physically insightful and capable of efficiently capturing the effects of disorder.

The Einstein model[15] and Cahill-Pohl model[16] provide clear and instructive physical pictures for understanding strongly scattered heat transport. The Einstein model treats atoms as independent harmonic oscillators with a single characteristic frequency, assuming energy is transferred only between neighboring atoms, with the phonon mean free path fixed at the interatomic distance. The Cahill-Pohl model extends this idea to a continuous vibrational spectrum and defines the phonon mean free path as half the wavelength ($\lambda/2$). Both models focus on heat transfer with transparent physical mechanisms and reasonable settings of phonon mean free path. Therefore, their core ideas are highly valuable and can be directly referenced to establish a reasonable thermal conductivity model for alloys.

SiGe and InGaAs alloys are selected as representative subjects owing to their extensive research attention and universal thermal transport evolution trends. Both SiGe and InGaAs exhibit superior carrier mobility, and thus are widely applied in advanced next-generation transistor architectures represented by gate-all-around (GAA) devices[17-19]. Besides, their composition-dependent thermal conductivity complies with common rules of alloys and follows the classic disorder-induced phonon scattering mechanism.

In this work, a substitutional atomic distance model (SADM) is proposed for predicting the thermal conductivity of alloys. By explicitly considering the spatial distribution of phonon scattering, this model delivers deeper physical insights into phonon transport behavior. First, the theoretical framework of the SADM is elaborated in detail. Specifically, the model is developed based on the VCA to acquire fundamental phonon properties, followed by the establishment of the alloy phonon scattering model. To demonstrate the model's performance and validate its feasibility, typical semiconductor alloys including SiGe and InGaAs are adopted for thermal conductivity calculation. The numerical results are comprehensively analyzed, which verifies the rationality and universal applicability of the proposed SADM.

## 2. Model

The SADM that explicitly accounts for the spatial distribution of disorder-induced scattering events was developed. In this framework, the phonon properties are estimated using the VCA introduced in Section 2.1, the alloy phonon scattering rates are determined by the SADM described in Section 2.2, and the lattice thermal conductivity is subsequently calculated as detailed in Section 2.3.

### 2.1 VCA for phonon properties

To compute the phonon properties required for the SADM, the lattice is treated using the VCA. In VCA, the disordered alloy is replaced by an effective "virtual" crystal whose atomic masses $M_{VCA}$, lattice parameters $M_{VCA}$, and force constants $F_{VCA}$ are compositionally averaged (as shown in Eqs. 1–3).

$$M_{VCA} = (1 - x) \cdot M_A + x \cdot M_B \tag{1}$$

$$a_{VCA} = (1-x) \cdot a_A + x \cdot a_B,\ a_{VCA} = (1-x) \cdot a_{AC} + x \cdot a_{BC} \quad (2)$$

$$F_{VCA} = (1-x) \cdot F_A + x \cdot F_B \quad (3)$$

Here, $x$ denotes the alloy composition, $m$ represents the atomic mass, and $a$ is the lattice constant. The subscripts $A$ and $B$ refer to different atomic species.

Using the virtual crystal parameters defined in Eqs. (1)–(3), the phonon properties of alloys such as $A_xB_{1-x}$ and $A_xB_{1-x}C$ were calculated. The dynamical matrix for the virtual crystal was constructed using the averaged masses and lattice constants, while the force constants $F_{VCA}$ were obtained by mixing the parameters from the pure A and B (or AC and BC) crystals, as defined in Eq. (3). This approach allows the calculation of phonon dispersion relations, density of states (DOS), group velocities, and anharmonic scattering rates within the VCA framework.

**2.2 Model for alloy scattering**

The SADM is shown in Fig. 1. Inspired by Einstein's minimum thermal conductivity, the interatomic distance is regarded as the phonon mean free path in Einstein's model. In this work, the phonon scattering mean free path induced by alloy disorder is determined by the average distance $d_{A-B}$ between atom *A* and its nearest neighbor atom *B*. This distance is considered to be the key factor leading to the deviation of the alloy structure from the VCA.

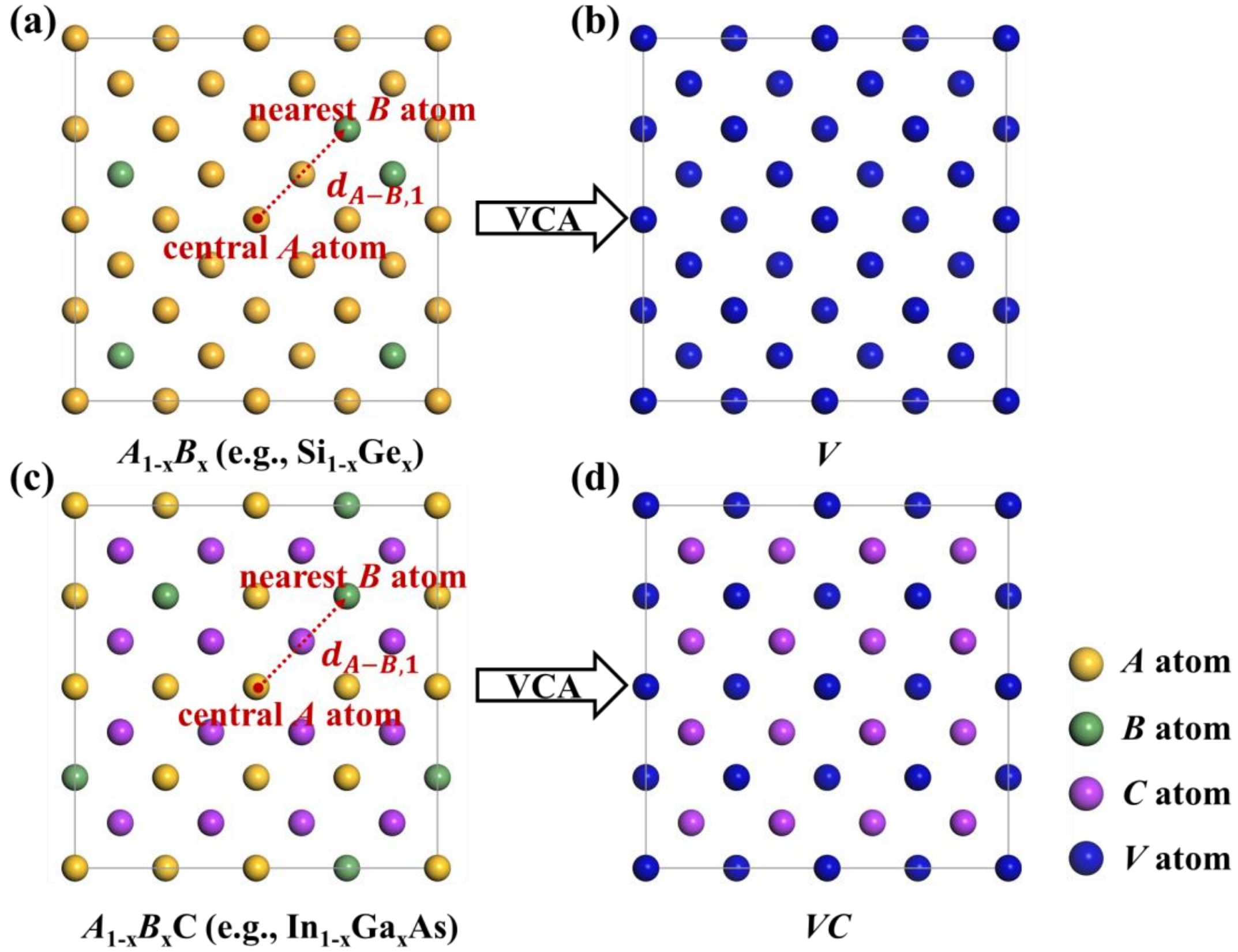

Fig. 1 Model schematic diagram for AB alloy (e.g., $Si_{1-x}Ge_x$) and ABC alloy (e.g., $In_{1-x}Ga_xAs$). The blue *V* atom represents the virtual atom after VCA.

The total phonon scattering rate can be obtained by applying Matthiessen's rule, as given in Eq. (4).

$$\tau_{total}^{-1} = \tau_{VCA,anharmonic}^{-1} + \tau_{alloy}^{-1} \tag{4}$$

The alloy phonon scattering rate is calculated as

$$\tau_{alloy}^{-1} = \Gamma \times \frac{v(\omega)}{d_{A-B}} \times \emptyset(\omega) \tag{5}$$

where $v(\omega)$ denotes the phonon group velocity, $d_{A-B}$ represents the average distance between scattering centers, The ratio $\frac{v(\omega)}{d_{A-B}}$ is regarded as the intrinsic scattering rate originating from alloy disorder. However, the actual scattering rate is affected by the alloy disorder parameter $\Gamma$（incorporating both mass disorder and force constant disorder）as well as the frequency-dependent term $\emptyset(\omega)$.

The disorder strength factor $\Gamma$ is given by

$$\Gamma = Cx(1-x) \times \left[ \left( \frac{\Delta M}{M_{VCA}} \right)^2 + \left( \frac{\Delta F}{F_{VCA}} \right)^2 \right] \tag{6}$$

where $x$ is the alloy composition, $\left(\frac{\Delta M}{M_{VCA}}\right)$ and $\left(\frac{\Delta F}{F_{VCA}}\right)$ are the differences in atomic mass and force constants, respectively, and $M_{VCA}$ and $F_{VCA}$ are the compositionally averaged mass and force constants, with all atoms within the system taken into account. The prefactor $C$ is a numerical constant calibrated via model fitting, which is set to 0.35 for both SiGe and InGaAs systems.

The average distance between scattering centers, $d_{A-B}$ is determined by the probability of finding a $B$ atom at the $i$-th site, as shown in Eq. (7) to (9).

$$p_{noneB}^{i} = (1-x)^{N_i} \tag{7}$$

$$p_B^i = \left( \prod_{j=1}^{i-1} p_{noneB}^{j} \right) \times \left(1 - p_{noneB}^{j}\right) \tag{8}$$

$$d_{A-B} = \sum_{i=1}^{\infty} p_B^i \times r_i \tag{9}$$

Here, $r_i$ is the distance to the $i$-th site, and $N_i$ denotes the number of atomic sites considered. This formulation accounts for the spatial distribution of scattering centers in the alloy.

The frequency-dependent factor $\emptyset(\omega)$ is given by

$$\emptyset(\omega) = \omega DOS(\omega) \tag{10}$$

where $DOS(\omega)$ is the phonon density of states. This term captures the reduced scattering for phonons with wavelengths much longer than the interatomic distance. In the low-frequency limit, this expression follows a $\omega^2 DOS(\omega)$ dependence of the scattering rate. Under the Debye approximation, where $DOS(\omega) \sim \omega^2$, this leads to a $\omega^4$ scattering law, which is consistent with previous studies.

## 2.3 Lattice thermal conductivity

The lattice thermal conductivity can be calculated using the phonon properties described in Section 2.1 and the phonon scattering rates obtained in Section 2.2, based on Eq. (11):

$$\kappa = \frac{1}{3}\int C(\omega)\, v(\omega)^2 \tau(\omega) d\omega \tag{11}$$

Where $C(\omega)$ is the spectral phonon heat capacity, $v(\omega)$ is the phonon group velocity, and $\tau(\omega)$ is the phonon relaxation time at frequency $\omega$.

# 3. Results and discussions

The phonon properties of SiGe and InGaAs semiconductor alloys were first calculated within the virtual crystal approximation, as described in Section 3.1. The alloy phonon scattering rates obtained from the SADM are presented in Section 3.2. The corresponding lattice thermal conductivities are discussed in Section 3.3.

## 3.1 Phonon properties

The phonon properties of SiGe and InGaAs alloys were calculated based on VCA. The second and third order force constants of Si, Ge, In, and GaAs were obtained from the almaBTE[20] database. Using Eqs. (1)–(3), these force constants were processed to construct the corresponding virtual crystals. The phonon dispersion curves were then calculated using Phonopy[21] package, as shown in Fig. 2. From the phonon dispersion, it can be observed that the cutoff frequency of Si is higher than that of Ge. As the alloy composition x increases (with increasing Ge content), the cutoff frequency gradually decreases and approaches that of Ge. Similar results can also be observed in InGaAs.

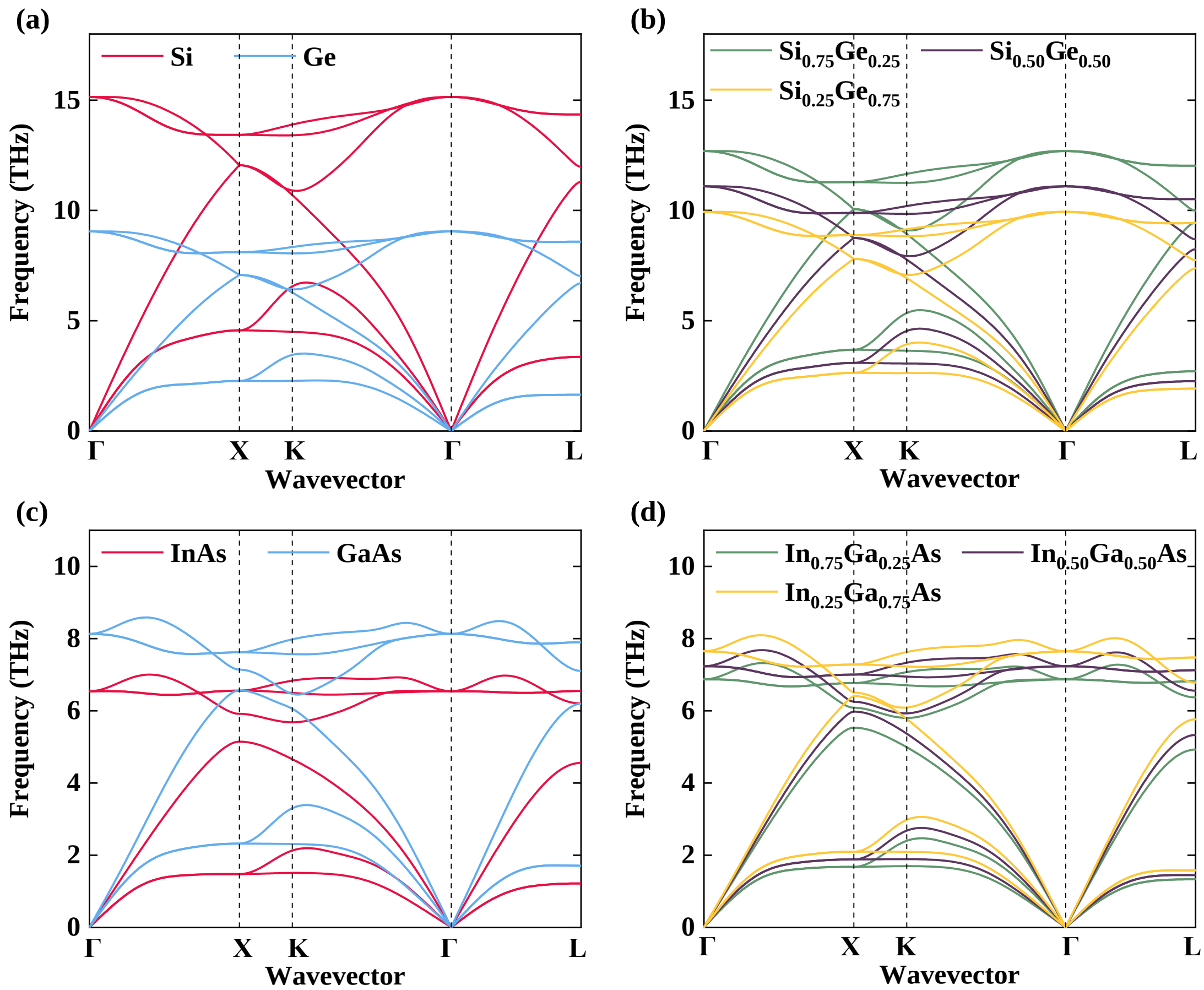

Fig. 2 Phonon dispersion relations of intrinsic semiconductors and their alloys. (a) Intrinsic Si (red) and Ge (blue); (b) $Si_{1-x}Ge_x$ alloys with (x=0.25, 0.50, 0.75) (green, purple, yellow, respectively); (c) Intrinsic InAs (red) and GaAs (blue); (d) $In_{1-x}Ga_xAs$ alloys with (x=0.25, 0.50, 0.75) (green, purple, yellow, respectively).

Subsequently, the DOS, group velocities and anharmonic phonon scattering rates were calculated. It is worth noting that, in computing the phonon group velocities, phonons along the entire Brillouin zone path were considered and averaged. The anharmonic phonon scattering rates were obtained using the ShengBTE[22] package, and detailed data of DOS, group velocities and anharmonic phonon scattering rates are provided in Supplementary Material (SM) I.

### 3.2 Phonon scattering rate

The alloy phonon scattering rates of SiGe and InGaAs calculated via the above method are displayed in Fig. 3. The vertical axis is plotted on a logarithmic scale, where the scattering rate spans approximately nine orders of magnitude and exhibits strong frequency dependence. The scattering rate rises sharply with increasing frequency in the low-frequency region and gradually saturates at medium and high frequencies. This variation tendency conforms to the typical frequency characteristics of phonon scattering rate reported in previous studies[9].

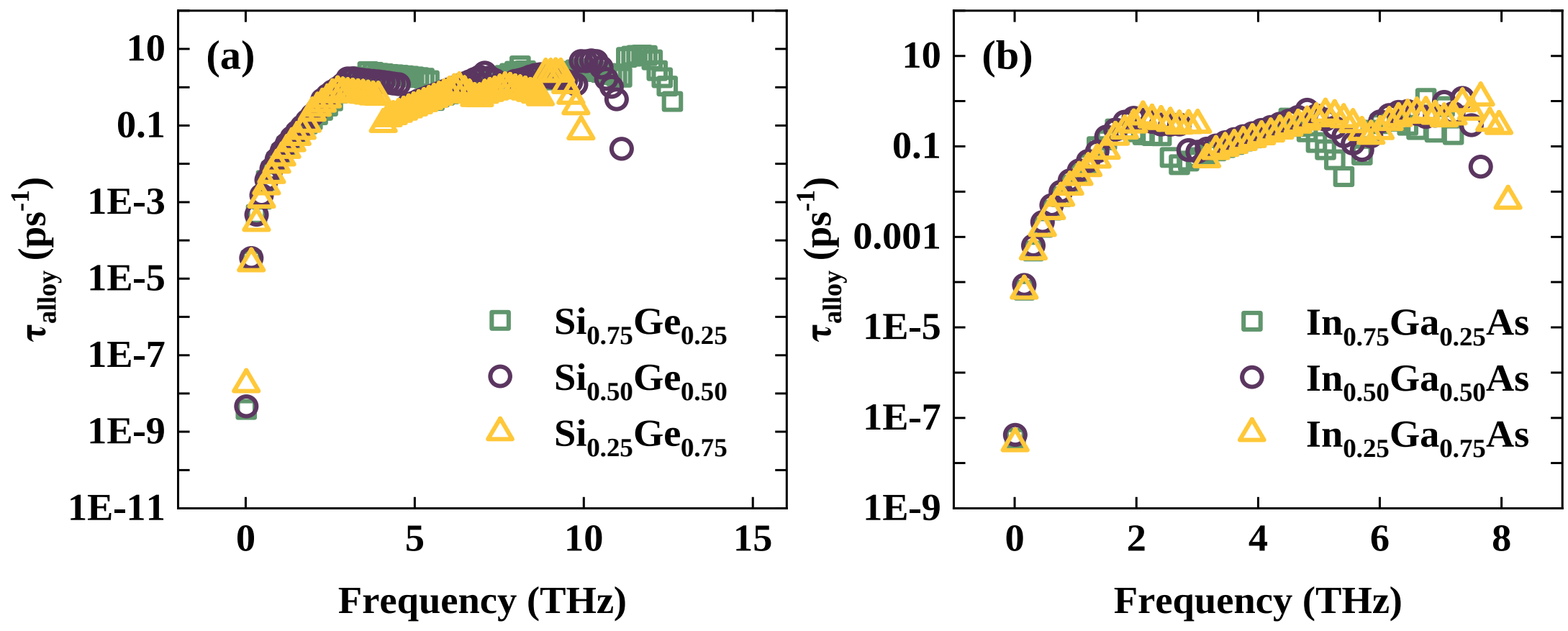

Fig.3 The alloy phonon scattering rates $\tau_{alloy}^{-1}$ for SiGe and InGaAs alloys at different compositions: (a) SiGe, (b) InGaAs

The total phonon scattering rates of SiGe and InGaAs are calculated and presented in Fig. 4. It can be seen that the phonon scattering rates of alloys are comparable to those of intrinsic semiconductors at frequencies below 1 THz. When the frequency exceeds 1 THz, the scattering rates of alloys increase remarkably. This demonstrates that alloy scattering mainly acts on phonons above 1 THz, which greatly weakens their contribution to thermal conductivity.

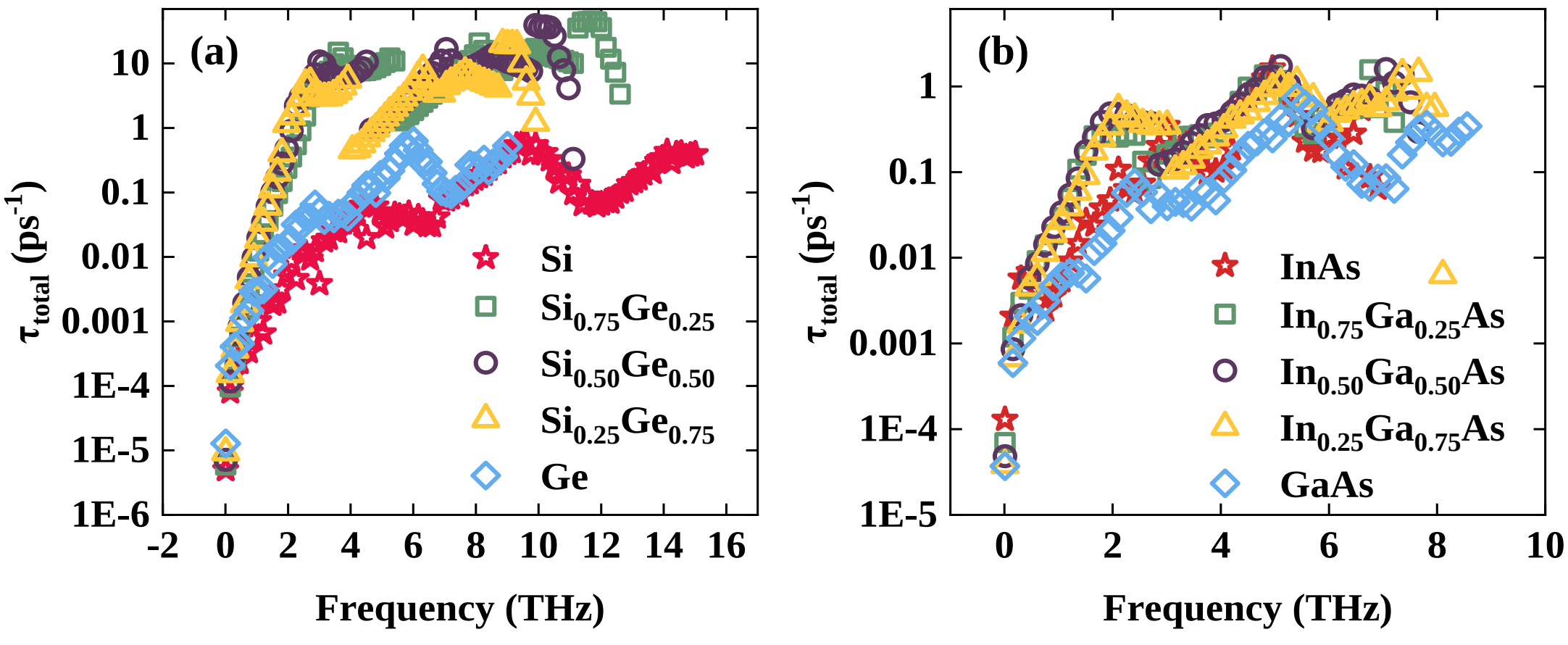

Fig.4 The total phonon scattering rates $\tau^{-1}$ for SiGe and InGaAs alloys at different compositions: (a) SiGe, (b) InGaAs.

### 3.3 Lattice thermal conductivity

The calculated compositional dependence of the lattice thermal conductivity for $Si_{1-x}Ge_x$ and $In_{1-x}Ga_xAs$ alloys are shown in Fig. 5, with compositional dependence with the vertical axis plotted on a logarithmic scale. Both alloys exhibit a characteristic U-shaped trend in thermal conductivity as a function of composition. As x varies from 0% or 100% toward intermediate values, the thermal conductivity decreases sharply, reaches a broad minimum plateau, and then gradually increases again. This universal U-curve originates primarily from alloy disorder scattering, as shown in Fig 3.

The predictions from this work are in good agreement with both published experimental measurements, verifying the reliability of the present model. The classical Tamura model delivers fairly consistent lattice thermal conductivity predictions[9] with experimental data for $Si_{1-x}Ge_x$ alloys, yet it yields substantial discrepancies between calculated[10] and measured thermal conductivities for $In_{1-x}Ga_xAs$ alloys. By contrast, calculations of SADM achieve excellent consistency with experimental values for both $Si_{1-x}Ge_x$ and $In_{1-x}Ga_xAs$ alloys. Meanwhile, the present results also closely match the thermal conductivity data reported by A et al. using the LDA-SQS combined T-matrix approach[10]. Collectively, these comparisons further validate the accuracy of our proposed model.

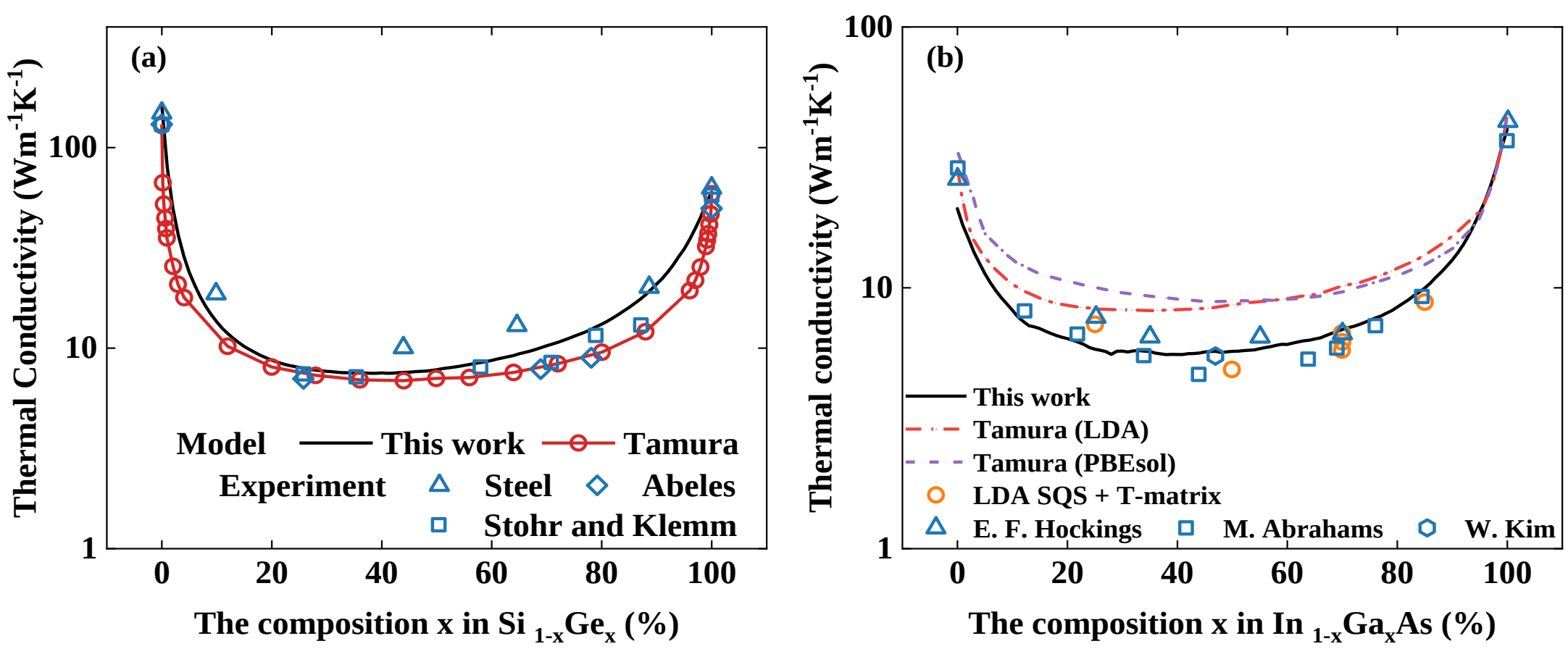

Fig. 5. Thermal conductivity of (a) $Si_{1-x}Ge_x$ and (b) $In_{1-x}Ga_xAs$ alloys as a function of composition x on a logarithmic scale. The black solid lines represent the results predicted in this work. (a) Red open circles are calculated results from the Tamura model[9]; The blue symbols denote experimental data: triangles[23], squares[9] and diamonds[8]. The relative error of alloy thermal conductivity between this work and experimental results is 17.64%. (b) The red and purple lines are the calculated results of the Tamura model using LDA and PBE-Sol pseudopotentials, respectively[10]. The yellow circles denote the predictions from the LDA-SQS + T-matrix method[10]. The blue symbols stand for experimental measurements, including triangles[24], squares[25] and hexagons[26]. The relative error of alloy thermal conductivity between this work and experimental results is 12.51%.

To analyze the contributions of phonons at different frequencies to the thermal conductivity, the cumulative thermal conductivity $\kappa_{cum}$ was calculated and is shown in Fig. 6. The cumulative thermal conductivity quantifies the contribution of phonons with frequencies $\omega$ to the total thermal conductivity. Compared with their intrinsic semiconductor counterparts, both $Si_{1-x}Ge_x$ and $In_{1-x}Ga_xAs$ alloys exhibit a much faster rise in normalized cumulative thermal conductivity at low frequencies. This is mainly because alloy scattering is much stronger for high-frequency phonons, significantly reducing their contribution to thermal conductivity.

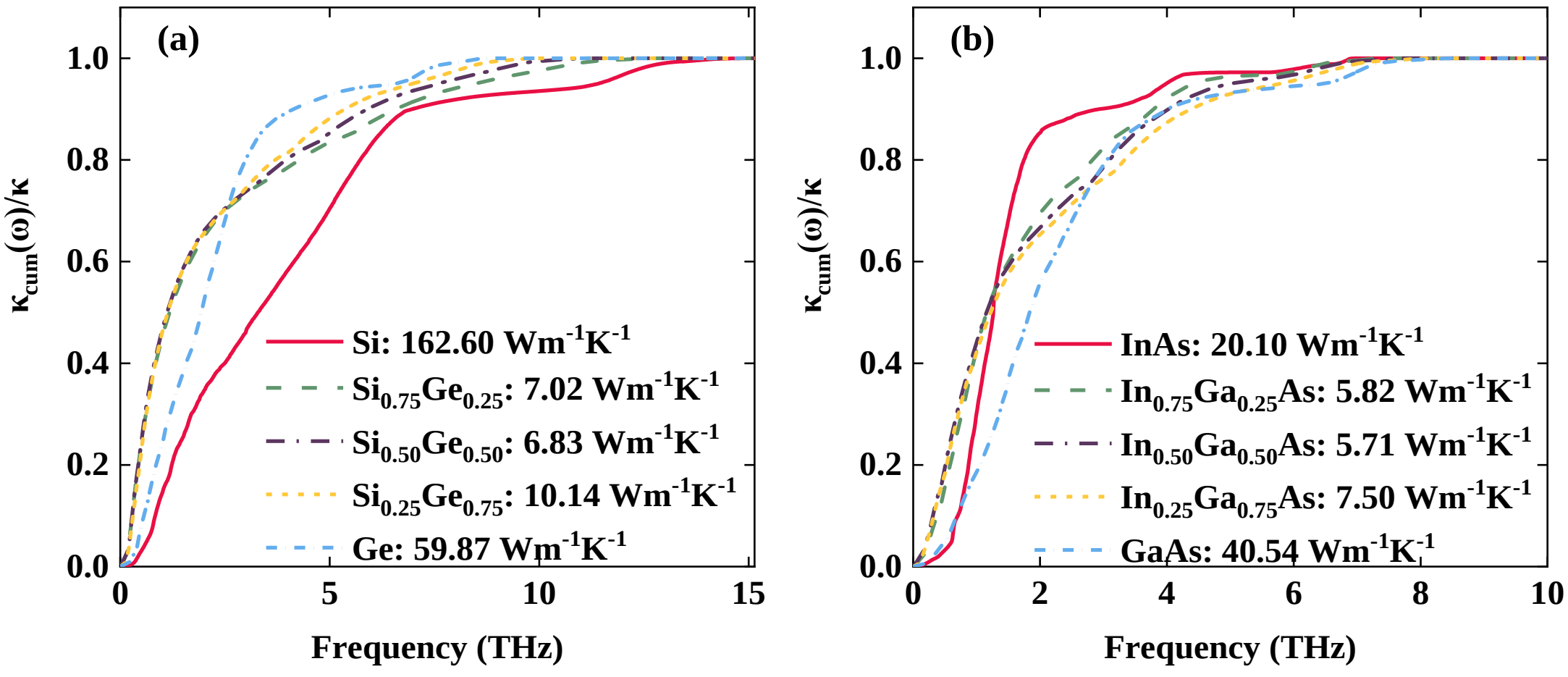

Fig.6 Cumulative thermal conductivity normalized to the total thermal conductivity plotted against phonon frequency for (a) $Si_{1-x}Ge_x$ and (b) $In_{1-x}Ga_xAs$ alloys. The value

in the legend indicates the thermal conductivity $\kappa$ for each composition.

## 4. Conclusion

In conclusion, this work proposed the SADM for predicting the thermal conductivity of semiconductor alloys. Distinct from conventional models, the developed SADM explicitly considers the spatial distribution of phonon scattering behaviors, thereby offering more in-depth and accurate physical insights into intrinsic phonon transport mechanisms within semiconductor alloy materials.

The theoretical framework of the SADM describes disorder-induced phonon scattering in alloys, with fundamental phonon parameters obtained via the VCA. Regarding alloy-induced phonon scattering, this model focuses on the spatial distribution characteristics of lattice disorder, namely the characteristic spatial scale, which is defined as the phonon mean free path dominating alloy scattering. Specifically, such spatial scale corresponds to the nearest-neighbor atomic distance between heterogeneous atoms in disordered sublattices of semiconductor alloys, which governs the average propagation distance of phonons between adjacent alloy scattering events and acts as the critical mean free path for alloy scattering. After acquiring phonon properties from VCA and determining corresponding phonon scattering rates, thermal conductivity can be efficiently calculated by the proposed model.

To validate the reliability and practicability of SADM, thermal conductivity simulations are conducted on two representative semiconductor alloys including SiGe and InGaAs. Comprehensive analysis of numerical simulation results further verifies the rationality and reliable applicability of the newly established SADM in predicting thermal conductivity of semiconductor alloys.

The present model provides a fundamental theoretical framework for further systematic investigations of semiconductor alloys. This work primarily focuses on the mutual substitution of two constituent elements in binary alloys, covering Si and Ge in SiGe alloys as well as In and Ga in InGaAs alloys. Future research can be further expanded and optimized in multiple dimensions, such as Si, Ge and Sn elements in SiGeSn alloys. Besides, subsequent studies can incorporate the configurational entropy of alloys to realize more accurate thermodynamic characterization of alloy systems, and further introduce the effects of short-range ordering and long-range ordering to perfect and enrich the theoretical system for alloy structural analysis.

## Declaration of competing interest

The authors declare that they have no known competing financial interests or personal relationships that could have appeared to influence the work reported in this paper.

## Acknowledgements

This work was sponsored by the Innovation Research Foundation of National University of Defense Technology (Innovation Research Foundation of NUDT). The work was carried out at the National Supercomputer Center in Tianjin, and the calculations were performed on TianHe-HPC.

## Data availability

Data will be made available on request.